\documentclass[conference,a4paper]{IEEEtran}

\usepackage{cite}
\usepackage{graphicx,color,epsfig,rotating}
\usepackage{amsfonts,amsmath,amssymb,bbm}
\usepackage{algorithm, algorithmic}
\usepackage{subfigure}
\usepackage{amsmath}
\usepackage{cite}
\usepackage{mdwtab}
\usepackage{subfigure}
\usepackage{placeins}
\usepackage{psfrag, graphicx}
\usepackage[latin1]{inputenc}
\usepackage{amssymb}
\usepackage{makeidx}
\usepackage{epstopdf}
\usepackage{url}

\long\def\comment#1{}

\newfont{\bbb}{msbm10 scaled 700}

\newfont{\bb}{msbm10 scaled 1100}

\newcommand{\eqdef}{\stackrel{\Delta}{=}}

\newcommand{\be}{\begin{equation}}
\newcommand{\ee}{\end{equation}}
\newcommand{\bea}{\begin{eqnarray}}
\newcommand{\eea}{\end{eqnarray}}

\newtheorem{theorem}{Theorem}

\begin{document}

\title{Cascaded Coded Distributed Computing on Heterogeneous Networks}

\author{Nicholas Woolsey, Rong-Rong Chen and Mingyue Ji
\thanks{The authors are with the Department of Electrical Engineering,
University of Utah, Salt Lake City, UT 84112, USA. (e-mail: nicholas.woolsey@utah.edu, rchen@ece.utah.edu and mingyue.ji@utah.edu)}
}

\author{
    \IEEEauthorblockN{ Nicholas Woolsey,
		Rong-Rong Chen, and Mingyue Ji }
	\IEEEauthorblockA{Department of Electrical and Computer Engineering, University of Utah\\
		Salt Lake City, UT, USA\\
		Email: \{nicholas.woolsey@utah.edu,
		 rchen@ece.utah.edu,
		mingyue.ji@utah.edu\}}

}

\maketitle

\thispagestyle{empty}
\pagestyle{empty}

\vspace{-0.5cm}

\begin{abstract}
Coded distributed computing (CDC) introduced by Li {\em et al.} in 2015 offers an efficient approach to trade computing power to reduce the communication load in general distributed computing frameworks such as MapReduce. For the more general \textit{cascaded} CDC, Map computations are repeated at $r$ nodes to significantly reduce the communication load among nodes tasked with computing $Q$ Reduce functions $s$ times. While an achievable cascaded CDC scheme was proposed, it only operates on homogeneous networks, where the storage, computation load and communication load of each computing node is the same. In this paper, we address this limitation by proposing a novel combinatorial design which operates on heterogeneous networks where nodes have varying storage and computing capabilities.
We provide an analytical characterization of the computation-communication trade-off and show that it is
optimal within a constant factor and could outperform the state-of-the-art homogeneous schemes. 
\end{abstract}



\section{Introduction}
\label{section: intro}
Coded distributed computing (CDC), introduced in \cite{li2018fundamental}, offers an efficient approach to reduce the communication load in  CDC networks such as MapReduce \cite{dean2008mapreduce}. 
In this type of distributed computing network, in order to compute the output functions, the computation is decomposed into ``Map" and ``Reduce" phases. First, each computing node computes intermediate values using  local input data files according to the designed Map functions. Then, computed intermediate values are exchanged among  computing nodes 
and nodes use these intermediate values as input to the designed Reduce functions to compute output functions.
The operation of exchanging intermediate values is called ``data shuffling" and occurs during the ``Shuffle" phase. This severely limits the performance of distributed computing applications due to the very high transmitted traffic load \cite{li2018fundamental}.

In \cite{li2018fundamental}, by formulating and characterizing a fundamental tradeoff between ``computation load" in the Map phase and ``communication load" in the Shuffle phase, Li  {\em et al.} demonstrated that these two quantities are inversely proportional to each other. This means that if each intermediate value is computed at $r$ carefully chosen nodes,
then the communication load in the Shuffle phase can be reduced by a factor of $r$. CDC achieves this multiplicative gain in the Shuffle phase by leveraging
coding opportunities created in the Map phase and strategically placing the input files among the computing nodes.
This idea was expanded on in \cite{konstantinidis2018leveraging}, \cite{woolsey2018new} where new CDC schemes were developed. However, a current limitation of these schemes is that they can only accommodate homogeneous computing networks. That is, the computing nodes have the same storage, computing and communication capabilities.

Understanding the performance potential and finding achievable designs for heterogeneous networks remains an open problem.
The authors in \cite{kiamari2017Globecom} derived a lower bound for the communication load for a CDC network where nodes have varying storage or computing capabilities. The proposed achievable scheme achieves the information-theoretical optimality of the minimum communication load for a system of $3$ nodes. The authors also demonstrated that the parameters of a heterogeneous CDC network can be translated into an optimization problem to find an efficient Map and Shuffle phase design.
In \cite{shakya2018distributed}, the authors studied CDC networks with $2$ and $3$ computing nodes where nodes have varying communication load constraints to find a lower bound on the minimum computation load.

In this paper, we focus on a specific type of CDC, called \textit{cascaded} CDC, where Reduce functions are computed at multiple nodes as opposed to just one node. According to our knowledge, other than \cite{li2018fundamental} and \cite{woolsey2018new}, the research efforts in CDC, including the aforementioned works, have focused on the case where each Reduce function is computed at exactly one node. However, in practice, 
it is often desired to compute each Reduce function multiple times. This allows for consecutive Map Reduce procedures as the Reduce function outputs can act as the input files for the next Map Reduce procedure \cite{zaharia2010spark}. 
\cite{li2018fundamental} proposed a cascaded CDC scheme to trade computing load for communication load. However, the achievable scheme only applies to homogeneous networks.

{\em Contributions:} In this paper, we propose a novel combinatorial design for cascaded CDC on heterogeneous networks where nodes have varying storage and computing capabilities. 
Meanwhile, the resulting computation-communication trade-off achieves the optimal trade-off within a constant factor 
for some system parameters.
Furthermore, compared to \cite{li2018fundamental}, our proposed design could achieve a better performance in terms of communication load while fixing other system parameters. It also greatly reduces the need for performing random linear combinations over intermediate values and hence, reduces the complexity of encoding and decoding in the Shuffle phase. To the best of our knowledge, this is the first work to explore heterogeneous cascaded CDC networks where Reduce functions are computed at multiple nodes. It offers the first general 
design architecture for heterogeneous CDC networks with a large number of computing nodes.
\vspace{-0.1cm}
\section{Network Model and Problem Formulation}
\label{sec: Network Model and Problem Formulation}

The network model is adopted from \cite{li2018fundamental}. We consider a distributed computing network where a set of $K$ nodes, labeled as $\{1, \ldots , K \}$, have the goal of computing $Q$ output functions and computing each function requires access to all $N$ input files. The  input files, $\{w_1 , \ldots , w_N \}$, are assumed to be of equal size of $B$ bits each. The set of $Q$ output functions is denoted by $\{ \phi_1 , \ldots \phi_Q\}$. Each node $k\in \{ 1, \ldots , K \} $ is assigned to compute a subset of output functions, denoted by $\mathcal{W}_k \subseteq \{ 1, \ldots Q \} $. The result of output function $i \in \{1, \ldots Q \}$ is $u_i = \phi_i \left( w_1, \ldots , w_N \right)$.

Alternatively, an output function can be computed  using ``Map" and ``Reduce" functions such that
\vspace{-0.1cm}
\be
u_i = h_i \left( g_{i,1}\left( w_1 \right), \ldots , g_{i,N}\left( w_N \right) \right),
\ee
where, for every output function $i$, there exists a set of $N$ Map functions $\{ g_{i,1}, \ldots , g_{i,N}\}$ and one Reduce function $h_i$. Furthermore, we define the output of the Map function, $v_{i,j}=g_{i,j}\left( w_j \right)$, as the \textit{intermediate value} resulting from performing the Map function for output function $i$ on file $w_j$. There are a total of $QN$ intermediate values  each with a size of $T$ bits.

The MapReduce distributed computing framework allows nodes to compute output functions without having access to all $N$ files. Instead, each node $k$ has access a subset of the $N$ files labeled as $\mathcal{M}_k \subseteq \{ w_1, \ldots , w_N\}$. We consider more general heterogeneous networks where the number of files stored at each nodes varies. Collectively, the nodes use the Map functions to compute every intermediate value in the {\em Map} phase at least once. Then, in the {\em Shuffle} phase, nodes multicast the computed intermediate values among one another via a shared link. 
The Shuffle phase is necessary so that each node can receive the necessary intermediate values that it could not compute itself. Finally, in the {\em Reduce} phase, nodes use the Reduce functions with the appropriate intermediate values as inputs to compute the assigned output functions. 
Throughout this paper, we consider cascaded CDC where each Reduce function is computed at $s$ nodes. 
However, different from \cite{li2018fundamental} where a Reduce function is assigned to any $s \in \{1, \cdots, K\}$, we will use a different function assignment based on a combinatorial design.

This distributed computing network design yields two important performance parameters: the computation load, $r$, and the communication load, $L$. The computation load is defined as the average number of times each intermediate value is computed among all  computing nodes. In other words, the computation load is the number of intermediate values computed in the Map phase normalized by the total number of unique intermediate values, $QN$. The communication load is defined as the amount of traffic load (in bits) among all the nodes in the Shuffle phase 
normalized by $QNT$. We define the computation-communication function as
\vspace{-0.1cm}
\be
L^*(r,s) \eqdef \inf\{L: (r, s, L) \text{ is feasible}\}.
\ee

\section{A $2$-dimensional Example}

\begin{figure}
\centering \includegraphics[width=9cm]{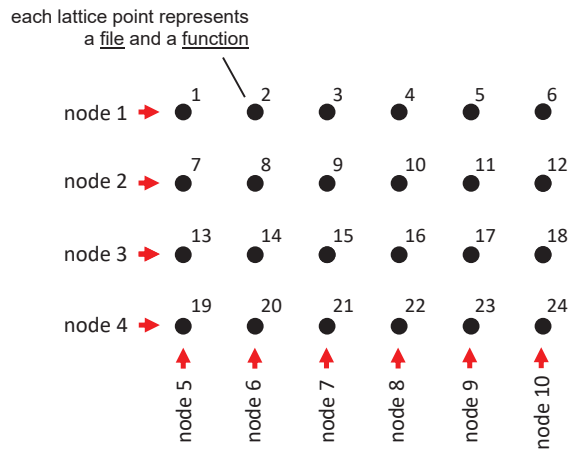}
\vspace{-0.5cm}
\caption{~\small A 2-dimensional lattice that defines the file availability and reduce function assignment amongst a heterogeneous CDC network of 10 computing nodes. Each lattice point represents a file and a function and nodes are assigned files and functions corresponding to a line of lattice points.}
\label{fig: het_2d_exmp}
\vspace{-0.4cm}
\end{figure}

We will illustrate the proposed design approach using an example, where we consider $K=10$ nodes, from which nodes $\{1,2,3,4\}$ have access to $\frac{6}{24}=\frac{1}{4}$ of the input files and nodes $\{5,6,7,8,9,10\}$ have access to $\frac{4}{24}=\frac{1}{6}$ of the input files. There are $Q=24$ functions and $N=24$ files. The specific file assignments are shown in Fig.~\ref{fig: het_2d_exmp}.  Two node sets are aligned along different dimensions of a rectangular lattice grid. Each lattice point represents both {\em a file} and {\em a Reduce function}.  The files and functions assigned to nodes are represented by a line of lattice points. For example, node $2$ has access to files of set $\mathcal{M}_2$ and assigned Reduce functions of set $\mathcal{W}_2$ where
$\mathcal{M}_2 = \mathcal{W}_2 =\{w_i  : i\in\{7 , \ldots , 12 \}\}$. 
Alternatively, node $8$ has access to files of set $\mathcal{M}_8$ and assigned Reduce functions of set $\mathcal{W}_8$ where
$\mathcal{M}_8 = \mathcal{W}_8= \{w_i\in\{4 , 10, 16 , 22\}\}$. 
Every input file is locally available at exactly $2$ nodes and therefore, $r=2$. Furthermore, every Reduce function is computed at $2$ nodes and $s=2$.

The intermediate values are classified by the number of nodes that request them in the Shuffle phase. Here we use ``request" to describe an intermediate value that a node needs to compute its assigned Reduce function, but cannot locally compute the intermediate value. In this example, intermediate values are requested by $0$, $1$ or $2$ nodes. For example, nodes $2$ and $8$ are both assigned Reduce function $10$, which requires $v_{10,10}$. However, nodes $2$ and $8$ have access to file $w_{10}$ and can locally compute $v_{10,10}$. Therefore, it is unnecessary to shuffle $v_{10,10}$ since no nodes request it. Reduce function $10$ also requires $v_{10,7}$ which node $2$ can compute locally, but node $8$ cannot. Because of this, $v_{10,7}$ is only requested by $1$ node. Finally,  $v_{10,2}$ is an example of an intermediate value requested by $2$ nodes, since neither node $2$ nor $8$ has access to file $w_2$.

The Shuffle phase consists of $2$ rounds. In the first round, intermediate values requested by $1$ node are shuffled. Nodes transmit coded intermediate value pairs to $2$ nodes aligned along the other dimension. For instance, node $1$ transmits $v_{1,2}\oplus v_{2,1}$ to nodes $\{ 5,6\}$ and $v_{4,6}\oplus v_{6,4}$ to nodes $\{ 8,10\}$.
Following this pattern, all intermediate values requested by $1$ node are transmitted in coded pairs.
 {In the first round, there are $4\cdot {6 \choose 2} + 6\cdot {4 \choose 2}=96$ transmissions, each of size  $T$ bits.}

In the second round, intermediate values requested by $2$ nodes are shuffled.
Consider every node set $\mathcal{S}$ of size 4 that includes  $2$ nodes from each dimension of the lattice grid. For instance, given the node set $\mathcal{S}=\{ 1,2,5,6\}$, we see that every intermediate value in $\{ v_{1,8}, v_{8,1}, v_{2,7}, v_{7,2} \}$ is requested by $2$ nodes in $\mathcal{S}$ and locally computed at the other $2$ nodes of $\mathcal{S}$.
Each intermediate value is split into $3$ non-overlapping packets and each node of $\mathcal{S}$ transmits $2$ linear combinations of its available packets. Hence, each node receives from the other 3 nodes in  $\mathcal{S}$  a total of $6$ linear combinations to solve for the $6$ unknown packets. This process is repeated for all possible choices of $\mathcal{S}$. { In the second round, we consider ${6 \choose 2}\cdot {4 \choose 2}=90$ node groups that each performs $4\cdot 2= 8$ transmissions of size $\frac{T}{3}$ bits. By counting all the transmitted bits over both rounds and normalizing by $QNT$, we find that $L = \frac{96+ 90\cdot 8 /3}{24\cdot 24}=\frac{7}{12} $.}

\section{General Achievable Scheme}
\label{sec: gen_sd}

In this section, we describe the general scheme to design a distributed computing network of $K$ nodes which collectively compute $Q$ functions $s$ times. Nodes are split into $s$ disjoint sets denoted by $\{\mathcal{K}_1,\ldots ,\mathcal{K}_s\}$ where $|\mathcal{K}_i|=x_i\geq 2$, $K = \sum_{i=1}^{s}x_i$. Also, $s\geq 2$ and  $s\in \mathbb{Z}^+$. To define file availability and  assign output functions, consider all node sets $\{\mathcal{T}:|\mathcal{T}\cap\mathcal{K}_i|=1\text{ }\forall \text{ } i\in \{1,\ldots , s\}\}$. There are $X = \prod_{i=1}^{s}x_i$ such sets which will be denoted as $\{\mathcal{T}_1,\ldots ,\mathcal{T}_{X}\}$. The $N$ input files and $Q$ output functions are both split into $X$ disjoint sets labeled as $\{\mathcal{B}_{1},\ldots,\mathcal{B}_{{X}}\}$ and $\{\mathcal{D}_{1},\ldots,\mathcal{D}_{{X}}\}$, respectively. Every node of the set $\mathcal{T}_i$ has access to the input files of $\mathcal{B}_{i}$ and is assigned the output functions of $\mathcal{D}_{i}$. The file sets are of size $\eta_1\in \mathbb{Z}^+$ where $N=X\eta_1$ and the output function sets are of size $\eta_2\in \mathbb{Z}^+$ where $Q=X\eta_2$. Furthermore, we define
$\mathcal{M}_k=\cup_{i : k\in \mathcal{T}_i}\mathcal{B}_i$
as the set of input files available to node $k$, and
$\mathcal{W}_k=\cup_{i : k\in \mathcal{T}_i}\mathcal{D}_i$
as the set of output functions assigned to node $k$.

In the following, we first define the Map phase. Then, the Shuffle phase contains $s$ rounds, where intermediate values requested by  $\gamma$ nodes are shuffled in the $\gamma$-th round. We propose two different methods to exchange intermediate values in the $\gamma$-th round.  Method A works for $1\leq \gamma \leq s-1$ and we consider groups of $2\gamma$ nodes. A node outside of a node group multicasts coded pairs of intermediate values to this node group. For Method B, nodes also form groups of $2\gamma$ nodes; however, nodes of this group multicast linear combinations of packets amongst one another.  Method B works for any round.

\begin{itemize}
\item {\bf Map Phase:} For all $k\in\{1,\ldots , K\}$, node $k$ computes intermediate value, $v_{i,j}$, if $i\in \{1, \ldots , Q \}$ and
$w_j\in \mathcal{M}_k$.

\item {\bf Shuffle Phase:} The Shuffle phase consists of $s$ rounds and
the $\gamma$-th round of the Shuffle phase is performed with one of the following methods:

\hspace{5mm}\textit{Method A ($1\leq \gamma \leq s-1$)}: Consider every node set $\mathcal{S}$ such that $|\mathcal{S}\cap \mathcal{K}_m|=2$ for all $m \in \mathcal{A} \subset \{ 1, \ldots, s\}$ where $|\mathcal{A}| = \gamma$ and $|\mathcal{S}|=2\gamma$. Furthermore, given $\mathcal{S}$ and $\mathcal{A}$, we consider every node set $\mathcal{Y}$ such that $|\mathcal{Y}\cap \mathcal{K}_m|=1$ for all $m \in\{1,\ldots , s \}\setminus \mathcal{A}$ and $|\mathcal{Y}|=s-\gamma$.
    For a set of nodes, $\mathcal{S}' \subset \mathcal{S}$ where $|\mathcal{S}'\cap \mathcal{K}_m|=1$ for all $m \in \mathcal{A} \subset \{ 1, \ldots, s\}$ and $|\mathcal{S}'|=\gamma$, define set $\mathcal{V}_{\mathcal{S}'\cup \mathcal{Y}}^{\mathcal{S}\setminus \mathcal{S}'}$ as the set of intermediate values requested only by nodes in  $\mathcal{S}\setminus \mathcal{S}'$ and computed only by nodes in $\mathcal{S}'\cup \mathcal{Y}$. More rigorously,
\begin{align}
&\mathcal{V}_{\mathcal{S}'\cup \mathcal{Y}}^{\mathcal{S}\setminus \mathcal{S}'}
 = \Big\{ v_{i,j} : i\in \mathcal{D}_\alpha, w_j \in \mathcal{B}_\ell , \nonumber \\
& \text{ }\text{ }\text{ }\text{ }\text{ }\text{ }\text{ }\text{ }\text{ }
\mathcal{T}_\alpha = \{\{\mathcal{S}\setminus \mathcal{S}'\}\cup \mathcal{Y}\}, \mathcal{T}_\ell =  \{\mathcal{S}'\cup\mathcal{Y}\}  \Big\}.
\end{align}
Similarly, we define
\begin{align}
&\mathcal{V}_{\{\mathcal{S}\setminus \mathcal{S}'\}\cup \mathcal{Y}}^{\mathcal{S}'}
\text{ }\text{ }= \Big\{ v_{i,j} : i\in \mathcal{D}_\alpha, w_j \in \mathcal{B}_\ell, \nonumber \\
& \text{ }\text{ }\text{ }\text{ }\text{ }\text{ }\text{ }\text{ }\text{ }
 \mathcal{T}_\alpha = \{\mathcal{S}'\cup\mathcal{Y}\} ,
 \mathcal{T}_\ell = \{\{\mathcal{S}\setminus \mathcal{S}'\}\cup \mathcal{Y}\}  \Big\}.
\end{align}
For every unique pair of $\mathcal{S}'$ and $\mathcal{S}\setminus\mathcal{S}'$, an arbitrary node in 
$\mathcal{Y}$ multicasts
$\mathcal{V}_{\mathcal{S}'\cup \mathcal{Y}}^{\mathcal{S}\setminus \mathcal{S}'} \oplus \mathcal{V}_{\{\mathcal{S}\setminus \mathcal{S}'\}\cup \mathcal{Y}}^{\mathcal{S}'}$
to the 
nodes of $\mathcal{S}$.

\hspace{5mm}\textit{Method B ($1 \leq \gamma \leq s$)}: Consider every node set $\mathcal{S}$ such that $|\mathcal{S}\cap \mathcal{K}_m|=2$ for all $m \in \mathcal{A} \subseteq \{ 1, \ldots, s\}$ where $|\mathcal{A}| = \gamma$ and $|\mathcal{S}|=2\gamma$. 
     Given sets $\mathcal{A}$ and $\mathcal{S}$, and for a set of nodes, $\mathcal{S}' \subset \mathcal{S}$ where
     $|\mathcal{S'}\cap \mathcal{K}_m|=1$ for all $m \in \mathcal{A} \subseteq \{ 1, \ldots, s\}$ and $|\mathcal{S}'|=\gamma$, define set $\mathcal{V}_{\mathcal{S}'}^{\mathcal{S}\setminus \mathcal{S}'}$ as the set of intermediate values requested only by nodes $\mathcal{S}\setminus \mathcal{S}'$ and computed by nodes $\mathcal{S}'$. More rigorously,
\begin{align}
\mathcal{V}_{\mathcal{S}'}^{\mathcal{S}\setminus \mathcal{S}'}  = \Big\{ v_{i,j} & : i\in \mathcal{D}_\alpha, w_j \in \mathcal{B}_\ell ,  \{\mathcal{S}\setminus \mathcal{S}'\} \subseteq \mathcal{T}_\alpha, \nonumber \\
&\quad\quad\quad\mathcal{S}' \subseteq \mathcal{T}_\ell, |\mathcal{T}_\alpha \cap \mathcal{T}_\ell|=s - \gamma  \Big\}
\end{align}
and this set of intermediate values is split into $2\gamma - 1$ equal size, disjoint subsets denoted by
$\left\{ \mathcal{V}_{\mathcal{S}'}^{\mathcal{S}\setminus \mathcal{S}'}[1], \ldots , \mathcal{V}_{\mathcal{S}'}^{\mathcal{S}\setminus \mathcal{S}'} [2\gamma -1]\right\}$. Every node $k\in \mathcal{S}$ multicasts $2^{(\gamma - 1)}$ linear combinations of the content of
$\left\{\mathcal{V}_{\mathcal{S}'}^{\mathcal{S}\setminus \mathcal{S}'} [a] : k\in \mathcal{S}', \text{ } a \in \{1, \ldots , 2\gamma - 1 \}\right\}$.
\item {\bf Reduce Phase:} For all $k\in \{ 1, \ldots , K \}$, node $k$ computes all output values $u_q$ such that $q\in \mathcal{W}_k$.
\end{itemize}

\section{A $3$-dimensional (cuboid) Example}

\begin{figure}
\centering \includegraphics[width=9cm]{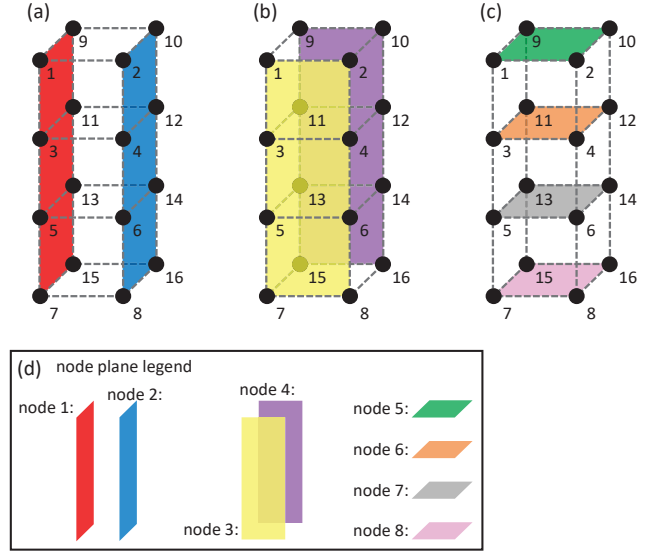}
\vspace{-0.65cm}
\caption{~\small A 3-dimensional lattice defining the file availability and reduce function assignment of 8 nodes in a heterogeneous CDC network. Each lattice point represents a file and a function. A node is represented by a plane in the lattice and its assigned files and functions are represented by the lattice points in the plane.}
\label{fig: het_exmp}
\vspace{-0.45cm}
\end{figure}

To illustrate the general scheme described in Section~\ref{sec: gen_sd}, we consider an example with $K=8$ computing nodes where
nodes $\{ 1,2,3,4\}$ have double the memory and computation power compared to nodes $\{ 5,6,7,8\}$. There are $X=16$ sets of files and functions where each file set contains just $1$ file ($\eta_1=1$) and each function set contains $1$ function ($\eta_2 = 1$) and $N=Q=16$. Files and functions are assigned to nodes represented by planes of a $3$-dimensional lattice grid as shown in Fig. \ref{fig: het_exmp}. The nodes are split into $3$ groups: $\mathcal{K}_1 = \{ 1,2\}$, $\mathcal{K}_2 = \{ 3,4\}$ and $\mathcal{K}_3 = \{ 5,6,7,8\}$, where nodes in each group are aligned along one  dimension of the lattice. Furthermore, as shown in  Fig. \ref{fig: het_exmp}, we define a node set $\mathcal{T}_i$ to contain  nodes whose file and function assignments, represented by a set of planes,  intersect at lattice point $i$. The nodes of $\mathcal{T}_i$ have the file $w_i$ locally available to them and are assigned Reduce function $i$.
In the Map phase, every node computes every intermediate value for each locally available file.

Next, we consider  the Shuffle phase. We adopt method A for the first two rounds and method B for the last round. In round $1$ ($\gamma=1$), we first consider pairs of nodes that are from the same set $\mathcal{K}_i$ and aligned along the same dimension. For instance, let $\mathcal{S} = \{ 1,2 \},  \mathcal{S'} = \{ 1\}$, and $\mathcal{Y}=\{3,8 \}$. We then have $\mathcal{S'} \cup \mathcal{Y}=\{1,3,8\}=\mathcal{T}_7$, 
and  $(\mathcal{S}\setminus \mathcal{S'}) \cup \mathcal{Y}=\{2,3,8\}=\mathcal{T}_8$.  Note that node $1$ is the only node that requests $v_{7,8}$ and node $2$ is the only node that requests $v_{8,7}$. Hence, either node $3$ or $8$ from $\mathcal{Y}$ can transmit $v_{7,8} \oplus v_{8,7}$ to nodes $1$ and $2$ in $\mathcal{S}$.  Note that the intermediate values requested by a single node are transmitted in coded pairs.
We continue this process by considering all possible choices of $\mathcal{S}, \mathcal{S'}, \mathcal{Y} $.   {In the first round, there are $ 2\cdot2\cdot{4 \choose 2} + 2~\cdot~4~\cdot~{2 \choose 2} + 2\cdot4\cdot{2 \choose 2} = 40$ transmissions, and each of size $T$ bits.}

Next, we consider round $2$ ($\gamma=2$) involving groups of $4$ nodes where $2$ are from $\mathcal{K}_i$ and $2$ are from $\mathcal{K}_j$, $i\neq j$. For instance, let $\mathcal{S}=\{3,4,6,8\}$.
 If we let $\mathcal{S'} = \{ 3,6\}$, and $\mathcal{Y}=\{1 \}$, we have $\mathcal{S'} \cup \mathcal{Y}=\{3,6,1\}=\mathcal{T}_3$, and  $(\mathcal{S}\setminus \mathcal{S'}) \cup \mathcal{Y}=\{4,8,1\}=\mathcal{T}_{15}$.
 Thus, node $1$ from $\mathcal{Y}$ will transmit $v_{3,15}\oplus v_{15,3}$ to $\mathcal{S}$.
 Note that intermediate values requested by $2$ nodes are also transmitted in coded pairs. { In the second round,  by considering all possible choices of $\mathcal{S}, \mathcal{S'}, \mathcal{Y} $, we see that there are $ 2\cdot2\cdot{2 \choose 2}\cdot{4 \choose 2} + 2\cdot2\cdot{2 \choose 2}\cdot{4 \choose 2} + 2\cdot4\cdot{2 \choose 2}\cdot{2 \choose 2} = 56$ transmissions, and each of size $T$ bits.}

 Finally, we consider round $3$ ($\gamma=3$) involving groups of $6$ nodes that contain $2$ nodes from each set $\mathcal{K}_1$, $\mathcal{K}_2$ and $\mathcal{K}_3$. For instance, consider $\mathcal{S} = \{ 1,2,3,4,5,6\}$. If we choose $\mathcal{S'} = \{ 1,3,5\} = \mathcal{T}_1$, then we have $\mathcal{S}\setminus \mathcal{S'}=\{2,4,6\}=\mathcal{T}_{12}$. We observe that $v_{1,12}$ is requested by three nodes in $\mathcal S'$ and is computed by all three nodes in   $\mathcal{S}\setminus \mathcal{S'}$.
 Similarly, by considering other choices of $\mathcal{S'} \subset  \mathcal{S}$,
   we identify
 8 intermediate values which are requested by 3 nodes of $\mathcal{S}$ and locally computed at the $3$ other nodes of $\mathcal{S}$. These  are: $v_{1,12}$, $v_{12,1}$,  $v_{3,10}$, $v_{10,3}$, $v_{4,9}$, $v_{9,4}$ and $v_{2,11}$, $v_{11,2}$. Each intermediate value is then split into $2 \gamma -1=2 \cdot 3 -1=5$ equal size packets and each node of $\mathcal{S}$ transmits $2^{\gamma-1}= 2^2=4$ linear combinations of its locally available packets. Each node collectively receives $4 \cdot 5=20$ linear combinations from the other $5$ nodes in $\mathcal S$ which are sufficient to solve for the requested $4$ intermediate values or $20$ unknown packets. { In the third round, we consider ${4 \choose 2}\cdot {2 \choose 2}\cdot {2 \choose 2} = 6$ node groups  that each performs $6\cdot 4= 24$ transmissions. Each transmission has a  size of  $\frac{T}{5}$ bits. This results in a total of $6 \cdot 6 \cdot \frac{4}{5}  =28.8$ normalized transmissions, each of size $T$, for round $3$.  By counting all the transmitted bits in three rounds and normalizing by $QNT$, we find $L =\frac{40+56+28.8}{256}= 0.4875$.}

\section{Achievable Computation and Communication Load}

The following theorem characterizes the computation and communication load for the heterogeneous design.

\begin{theorem}
\label{theorem: 1}
Let $K,Q,N,s$ be the number of nodes, number of functions, number of files, and number of nodes which compute each function, respectively. For some $x_1,\ldots,x_s\in \mathbb{Z}^+$ where $X = \prod_{i=1}^{s}x_i$ and $s,\eta_1,\eta_2 \in \mathbb{Z}^+$ such that $K=\sum_{i=1}^{s}x_i$, $s\geq 2$, $N = \eta_1 X $ and $Q = \eta_2 X $, the following computation and communication load pair is achievable:
\begin{eqnarray}
\label{eq: eqL_sgt1}
r_{\rm c} &=& s \\
L_{\rm c}(r_{\rm c}, s) &=& \frac{1}{2X}\sum_{\gamma = 1}^{s-1}\left[\sum_{\substack{ \{\mathcal{A}:\mathcal{A}\subset \left[s\right],  |\mathcal{A}|=\gamma\}}}\left[\prod_{i\in\mathcal{A}}\left( x_i-1\right)\right]\right] \nonumber \\
&&\quad +\frac{s}{X\left(2s-1 \right)}\prod_{i=1}^{s}\left(x_i-1\right) \label{eq: L_het}
\end{eqnarray}
where $[s]=\{ 1,\ldots , s\}$. Theorem \ref{theorem: 1} is proved in Appendix \ref{sec: pf_th_1}.
\end{theorem}

Given the file and function assignment of our scheme, the optimality of $L_{\rm c}$ is given by the following theorem. 
\begin{theorem}
\label{th: conv}
  Given the function and file assignment defined by the cascaded CDC design  in Section \ref{sec: gen_sd},  
  let $L_{\rm c}^*(r_{\rm c},s)$ be the optimal 
  communication load over all achievable Shuffle schemes; $L_c(r_{\rm c},s)$ is given in (\ref{eq: L_het}) and $r_c$ is given in (\ref{eq: eqL_sgt1}), then 
  \be
  1 \leq \frac{L_{\rm c}(r_{\rm c}, s)}{L_{\rm c}^*(r_{\rm c},s)} \leq 4.
  \ee
 \hfill$\square$
\end{theorem}
%

\section{Simulations}

\label{sec: disc_sg1}

We perform a simulation to compare the new heterogeneous network design to the homogeneous designs, either of \cite{li2018fundamental} or of our previously proposed combinatorial design in \cite{woolsey2018new}. We focus on the storage and computation capabilities of nodes in a network, where some nodes have more storage than other nodes such that they can also compute more intermediate values of all output functions. In comparing all the schemes we fix $r=s=3$. We vary $K$, and for each value of $K$, we keep $N$ and $Q$ constant across the schemes.\footnote{
We adjust $N$ and $Q$ by using the appropriate $\eta_1$ and $\eta_2$ coefficients.} Furthermore, for the heterogeneous design $\frac{1}{3}$ of the nodes have $4$ times as much storage capacity and computing power compared to the other $\frac{2}{3}$ of the nodes.

\begin{figure}
\centering
\centering \includegraphics[width=8.0cm]{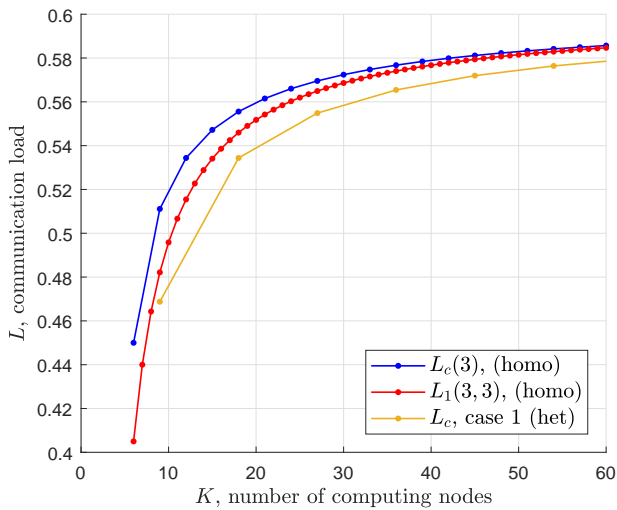} 
\vspace{-0.2cm}
\caption{~\small A comparison of the proposed heterogeneous cascaded CDC design and corresponding homogeneous CDC designs. 
}
\label{fig: het sim3}
\vspace{-0.4cm}
\end{figure}

{ The simulation results are shown in Fig.~\ref{fig: het sim3}, where $L_{\rm c}(3)$ (homo) and $L_1(3,3)$ (homo) refer to the homogeneous schemes in \cite{li2018fundamental} and \cite{woolsey2018new}, respectively, and $L_{\rm c}$ (het) refers to the proposed heterogeneous scheme.} 
Interestingly, the communication load of the new heterogeneous design is lower than those of the  state-of-the-art homogeneous designs, possibly due to the advantage of having a set of nodes with a greater number of  locally available files and assigned functions than other nodes. In this way, less shuffling is required to satisfy the requests of these nodes. An extreme case of this can be observed when a subset of nodes each has local access to all files and collectively compute all functions. In this case, 
the communication load is $0$. 

\section{Conclusions}

In this work, we  proposed a novel combinatorial design for heterogeneous cascaded CDC networks where nodes have varying storage and computing capabilities. The proposed architecture allows reduce functions to be computed at multiple nodes, and hence is amicable for practical implementations of MapReduce systems.
An analytical characterization of the computation-communication trade-off  shows that the proposed design is
 optimal within a constant factor and could outperform the state-of-the-art homogeneous schemes. This work reveals that it is advantageous to explore heterogeneous CDC systems due to reduced communication load that comes with nodes with larger memory and computing power.
\appendices

\section{Proof of Theorem \ref{theorem: 1}}
\label{sec: pf_th_1}
Every file is assigned to $s$ nodes and every node computes all intermediate values from its locally available files. Therefore, a total of $sNQ$ intermediate values are calculated. After normalizing by $QN$, we obtain that $r_{\rm c} = s$.

The communication load can be calculated by considering all $s$ rounds of the Shuffle phase. In the $\gamma$-th round, we consider a set $\mathcal{S}$ of $2\gamma$ nodes where there are 2 nodes from $\mathcal{K}_i$ for all $i \in \mathcal{A}\subseteq \{1, \ldots s \}$ such that $|\mathcal{A}|=\gamma$.  Given $\mathcal{A}$ and $\mathcal{S}$ we identify all node sets $\mathcal{Y}$ which contain $s-\gamma$ nodes, $1$ node from each set $\mathcal{K}_i$ for all $i \in \{ 1, \ldots , s\}\setminus \mathcal{A}$. Given $\mathcal{A}$,  there are
$\prod_{i\in\mathcal{A}}{x_i \choose 2}$
possibilities for a group $\mathcal{S}$, and
$\prod_{i\in [s]\setminus \mathcal{A}}x_i$
 possibilities for $\mathcal{Y}$.
Furthermore, there are $2^\gamma$ possibilities for a subset $\mathcal{S}'\subset \mathcal{S}$ such that $|\mathcal{S}'|=\gamma$ and all nodes of $\mathcal{S}'$ share a common request because $\mathcal{S}'$ must contain 1 node from each set $\mathcal{K}_i$ for all $i\in\mathcal{A}$ in order for these nodes to share a common request. Therefore, there are
$2^\gamma\prod_{i\in\mathcal{A}}{x_i \choose 2}\prod_{i \notin \mathcal{A}}x_i= X \prod_{i\in\mathcal{A}} (x_i-1)$
unique pairs of $\mathcal{Y}$ and $\mathcal{S}'$ given $\mathcal{A}$. For each unique pair of $\mathcal{Y}$ and $\mathcal{S}'$, we define a set of intermediate values $\mathcal{V}_{\mathcal{S}'\cup \mathcal{Y}}^{\mathcal{S}\setminus \mathcal{S}'}$ which only contains intermediate values $v_{i,j}$ such that $i\in\mathcal{D}_\alpha$ and $w_j\in\mathcal{B}_\ell$ where $\{\{ \mathcal{S}\setminus\mathcal{S}'\}\cup \mathcal{Y} \}=\mathcal{T}_\alpha$ and $\mathcal{S}' \cup \mathcal{Y} = \mathcal{T}_\ell$. Since $|\mathcal{B}_\ell|=\eta_1$ and $|\mathcal{D}_\alpha|=\eta_2$, we see that $|\mathcal{V}_{\mathcal{S}'\cup \mathcal{Y}}^{\mathcal{S}\setminus \mathcal{S}'}|=\eta_1 \eta_2$. All of the intermediate value sets are transmitted in coded pairs, effectively reducing the contribution to the communication load by half. Therefore, given $\mathcal{A}$, there are
$\frac{\eta_1\eta_2X}{2} \prod_{i\in\mathcal{A}} (x_i-1)$
transmissions of size $T$ bits, the number of bits in a single intermediate value.
{  Accounting for all possibilities of $\mathcal{A}$ for $\gamma = 1, \ldots , s-1$, we obtain }
$\frac{\eta_1\eta_2}{2X}\sum_{\gamma = 1}^{s-1}\left[\sum_{\substack{ \{\mathcal{A}:\mathcal{A}\subset \left[s\right],  |\mathcal{A}|=\gamma\}}}\left[\prod_{i\in\mathcal{A}}\left( x_i-1\right)\right]\right]$
transmissions of size $T$ bits.

Finally, in the $s$-th round, we consider all possible groups of $2s$ nodes, $\mathcal{S}$, such that $|\mathcal{S}\cap \mathcal{K}_i|=2$ for all $i\in \{1, \ldots, s \}$. There are
$\prod_{i=1}^{s}{x_i \choose 2}$
possibilities for a group $\mathcal{S}$. Furthermore, given $\mathcal{S}$, there are $2^s$ possibilities for a group $S'\subset\mathcal{S}$ such that $|\mathcal{S}'\cap \mathcal{K}_i|=1$ for all $i\in \{1, \ldots, s \}$. We see that $\mathcal{S}'=\mathcal{T}_\ell$ and $\{ \mathcal{S}\setminus\mathcal{S}' \}=\mathcal{T}_\alpha$ for some $\ell$ and $\alpha$. Therefore, $|\mathcal{V}_{\mathcal{S}'}^{\mathcal{S}\setminus \mathcal{S}'}|=|\mathcal{B}_\ell|\cdot|\mathcal{D}_\alpha|=\eta_1 \eta_2$. Each node of $\mathcal{S}$ transmits $2^{s-1}$ linear combinations of size
$\frac{\eta_1\eta_2T}{2s-1}$
bits and the total number of bits transmitted in the $s$-th round is
\be
\frac{2s\eta_1\eta_22^{s-1}T}{2s-1}\prod_{i=1}^{s}{x_i \choose 2} = \frac{s\eta_1\eta_2TX}{2s-1}\prod_{i=1}^{s}(x_i -1).
\ee
By summing the number of bits transmitted over all rounds and normalizing by $QNT=\eta_1\eta_2X^2T$, we obtain (\ref{eq: L_het}).

\bibliographystyle{IEEEbib}
\bibliography{references_d2d_v2}

\end{document}